\documentclass[showpacs,pra,aps,nopacs,onecolumn,twoside,superscriptaddress]{revtex4}

\usepackage{amsmath,amsfonts,amssymb,caption,color,epsfig,graphics,graphicx,hyperref,latexsym,mathrsfs,revsymb,theorem,url,verbatim,epstopdf}

\hypersetup{colorlinks,linkcolor={blue},citecolor={red},urlcolor={blue}}
\usepackage{subfigure}
\pdfoutput=1

\def\squareforqed{\hbox{\rlap{$\sqcap$}$\sqcup$}}
\def\qed{\ifmmode\squareforqed\else{\unskip\nobreak\hfil
\penalty50\hskip1em\null\nobreak\hfil\squareforqed
\parfillskip=0pt\finalhyphendemerits=0\endgraf}\fi}
\def\endenv{\ifmmode\;\else{\unskip\nobreak\hfil
\penalty50\hskip1em\null\nobreak\hfil\;
\parfillskip=0pt\finalhyphendemerits=0\endgraf}\fi}

\def\Dbar{\leavevmode\lower.6ex\hbox to 0pt
{\hskip-.23ex\accent"16\hss}D}
\makeatletter
\def\url@leostyle{%
  \@ifundefined{selectfont}{\def\UrlFont{\sf}}{\def\UrlFont{\small\ttfamily}}}
\makeatother
\newcommand{\bra}[1]{\langle#1|}
\newcommand{\ket}[1]{|#1\rangle}

\newcommand{\braket}[2]{\langle#1|#2\rangle}

\def\bma{\begin{bmatrix}}
\def\ema{\end{bmatrix}}
\def\Dbar{\leavevmode\lower.6ex\hbox to 0pt
{\hskip-.23ex\accent"16\hss}D}

\begin{document}\large

\title{Realization of controlled Remote implementation of operation}

\date{\today}

\author{Shaomin Liu}
\affiliation{School of Mathematics, Physics and Finance, Anhui Polytechnic University, Wuhu 241000, China}

\author{Qi-Lin Zhang}
\affiliation{School of Mathematics, Physics and Finance, Anhui Polytechnic University, Wuhu 241000, China}
\author{Lin Chen}\email[]{linchen@buaa.edu.cn (corresponding author)}
\affiliation{LMIB(Beihang University), Ministry of Education, and School of Mathematical Sciences, Beihang University, Beijing 100191, China}
\affiliation{International Research Institute for Multidisciplinary Science, Beihang University, Beijing 100191, China}

\begin{abstract}
Controlled remote implementation of operation (CRIO) enables to implement operations on a remote state with strong security. We transmit implementations by entangling qubits in photon-cavity-atom system. The photons transferring in fibre and the atoms embedded in optical cavity construct CZ gates. The gates transfer implementations between participants with the permission of controller.  We also construct nonadiabatic holonomic controlled gate between alkali metal atoms.  Decoherence and dissipation decrease the fidelity of the implementation operators. We apply anti-blockade effect and dynamical scheme to improve the robustness of the gate.
\end{abstract}
\maketitle
\section{Introduction}
\label{sec:int}
Quantum information enables to transmit information in long distance with strong security. Some groups realized quantum key distribution (QKD) over several hundreds or thousands kilometers via fibre-based \cite{6} and satellite-based \cite{7} transmission.
 Quantum communication includes transferring states for distance and implementing operations on a remote state. The non-local nature of entanglement enables to realize both of them experimentally.
In \cite{18,19,20,21}, authors suggested the implementation solutions of CRIO. They are useful in operation sharing, remote state preparation and distributed quantum computation.
Decoherence and dissipation obstruct the realization of existing CRIO protocols.

In \cite{4}, authors proposed CRIO protocols to realize RIO with the permission of the controller. In which, the quantum qubits belonging to different participants are entangled using the graph state.
With the help of the stator, the participants transmit local implementations to remote implementations \cite{5}.
To realize the protocols experimentally, there are two difficulties, preparing entanglement states in long distance  and realizing robust local controlled gates.

We solve the first problem by constructing photon-cavity-atom system. Photon is the best candidate of flying qubit because it can be transmitted in optical fibers effectively \cite{25}. The neutral atom in an optical cavity is chosen as the static qubit based on cavity quantum electrodynamics (QED). We construct basic logic qubit gates by coupling a $\Lambda$-type three level atom with one-sided optical cavity \cite{22,11}. Via the optical fiber, the information exchanges between the flying and static qubit. Due to the chromatic and polarization mode dispersion, the polarized state of photon is unstable
when it is transferred in fiber. We can choose polarization-maintaining (PM) fiber to protect the polarizer qubits.

For the second problem, Rabi oscillation of atom excited by photon field can realize universal gate easily. Because of the decoherence and control errors, we need more robust protocol.
 Nonadiabatic holonomic quantum computation (NHQC) \cite{23,30,29,34} is based on nonadiabatic non-Abelian geometric phases. It possesses whole geometric property. The geometric phases can be used as parameters of quantum gate. They are independent of the evolution details and robust against stochastic noises. NHQC has been realized in many physical systems, such as superconducting qubits \cite{37}, trapped ions \cite{38}, and semiconductor quantum dots \cite{39}. In this paper, we choose cold Rydberg atom, i.e. Rb, Cs, Sr..., as the platform to realize nonadiabatic holonomic controlled gate \cite{26,24,27,41}.
 Suppose there are two alkali metal atoms with distance less than $10\mu m$. The strong Coulomb interaction between atoms in the Rydberg state causes a Stark shift of the excited state. Two atoms can not be excited simultaneously. That is the Rydberg blockade effect \cite{31}. This effect is useful in constructing controlled gate. Since atom is excited in this scheme, the atomic decay decreases the fidelity of the gate.
 Anti-blockade scheme and dynamical gate \cite{28} are suggested to weaken the impact of decoherence factors.

The rest of this paper is constructed as follows. In Sec. \ref{sec:pre}, we introduce the protocol of CRIO proposed in \cite{4}. In Sec. \ref{sec:imp}, we suggest a set of schemes to realize CRIO by light fiber, cavity and optical dipole trap (ODT).  Sec. \ref{sec:fid} discusses the fidelity and efficiency of the protocol.
Sec. \ref{sec:con} makes conclusion and outlook.

\section{Preliminaries}
\label{sec:pre}
In this section, we introduce the CRIO protocol proposed in \cite{4}, and then discuss how to realize it experimentally.

\begin{figure}[h!]
\centerline{
\includegraphics[width=0.5\textwidth]{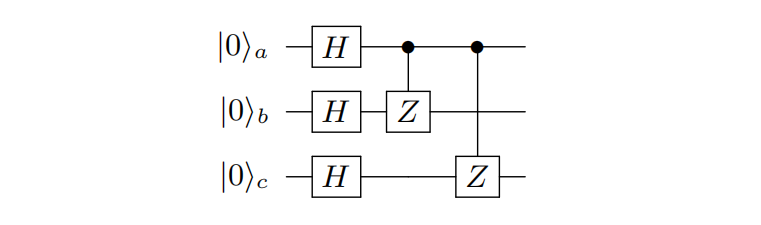}
}\caption{Preparation of tripartite graph state $\ket{h_3}$ \cite{4}.}
\label{fig:h3state}
\end{figure}
To realize CRIO, the first step is preparing the graph state.
In tripartite case, the control qubit $a$ belongs to Alice and the target qubits $b,c$ belong to Bob and Charlie respectively. As quantum circuits shown in Fig. \ref{fig:h3state}, the CZ gates between $a,b$ and $a,c$ generate the entangled state,
\begin{equation}
\label{eq:n=h3}
\begin{aligned}
  \ket{h_3} & =CZ_{(a,b)}CZ_{(a,c)}\ket{+}^{\otimes 3}\\
   & =\frac{1}{2\sqrt{2}}\big(\ket{000}+\ket{001}+\ket{010}+\ket{011}+\ket{100}-\ket{101}-\ket{110}+\ket{111}\big)_{a,b,c}.
\end{aligned}
\end{equation}
Then Charlie entangles another qubit $C$ in unknown state $\ket{\Psi}_C$ with $c$,
\begin{equation}
\label{eq:n=UcC}
U_{cC}=\ket{0}_c\bra{0}\otimes I_C+\ket{1}_c\bra{1}\otimes \sigma_{n_C}.
\end{equation}
The operator $\sigma_{n_C}=\vec{n_C}\cdot\vec{\sigma}$ satisfies $\sigma_{n_C}^2=I$, where $\vec{n}=( \sin\theta\cos\varphi,\sin\theta\sin\varphi,\cos\theta )$ and $\vec{\sigma}$ is the pauli matrix vector. This operator is known by Charlie only.

When these preparations are finished,
with the permission of Alice, Bob can implement a remote operation on Charlie's qubit $C$. C is in an unknown state and far away from Bob.
 The process is as follows.  Alice performs a measurement $\sigma_x$ on qubit $a$. If the result is $\ket{-}_a$, she informs Bob of the result and Bob performs the operation $\sigma_x$ on qubit $b$, otherwise Bob does nothing.
Then Charlie measures $\sigma_x$ on qubit $c$ and informs Bob of the result. If the result is $\ket{-}_c$, Bob implements the operation $\sigma_z$ on qubit $b$, otherwise Bob does nothing. The stator, a hybrid state operator, $S_3=\ket{0}_b\otimes I_C+\ket{1}_b\otimes \sigma_{n_C}$ is obtained between Bob and Charlie \cite{5}.
 Suppose Bob implements the operation
$e^{i\alpha\sigma_{x_b}}=\cos\alpha I_b+i\sin\alpha\sigma_{x_b} $ on qubit $b$, the real number $\alpha$ is determined by Bob only and unknown to other participants.  Bob measures qubit $b$ in the Z-basis and informs Charlie of the result.
If the result is $\ket{1}_b$, Charlie performs the local rotation $e^{i\pi\sigma_{n_C}/2}$ on qubit $C$. With the help of the eigenoperator equation $e^{i\alpha\sigma_{x_b}}S_3=e^{i\alpha\sigma_{n_C}}S_3$, the remote operation between Bob and Charlie is realized.
In the whole procedure, other participants know nothing about $\alpha$ except for Bob. The CRIO protocol presents strong security.

 In multipartite case, the remote operation passages are extended. For example, in five-partite case, the stator $S_5$ accomplishes the remote operation between qubit $b,D$ and $c,E$.
  Bob and Charlie implement local operations $e^{i\alpha\sigma_{x_b}}$ and $e^{i\beta\sigma_{x_c}}$ on their qubits $b$ and $c$ respectively.  The remote operations $e^{i\alpha\sigma_{n_D}}$ and $e^{i\beta\sigma_{n_E}}$ are implemented on qubit $D$ and $E$ via $\ket{h_5}$ simultaneously. This remote operation is implemented under the supervision of Alice.
  In $(2N+1)$-partite case, Alice also is the controller, determines whether and when these operations be implemented. There are N real numbers $\beta_j$ can be transmitted by the stator $S_{2N+1}$ from qubit $a_j$ to $a_{N+j}$ respectively, $j= 2, 3,..., N+1$.
So the protocol can be extended to multipartite remote operations well.

In this protocol, most implementations can be realized by local basic operations on qubits. The difficult part is how to realize long distance entanglement between participants Alice, Bob and Charlie as Eq. (\ref{eq:n=h3}) shown. We choose optical fibre to transfer the remote operation for its stability. Another problem we will discuss in next section is how to construct robust and efficiency controlled gate between cC in Eq. (\ref{eq:n=UcC}). The double partite controlled gate $\vec{n_C}\cdot\vec{\sigma}$ can be realized in many schemes, such as ion trap and Rydberg blockade. The latter one has advantages in individual addressing, extendibility and high gate speed.

\section{Realization of CRIO}
\label{sec:imp}
In this section, we discuss how to realize CRIO protocol in tripartite case. There are two problems, preparing $\ket{h_3}$ in Eq. (\ref{eq:n=h3}) and constructing the local controlled gate $U_{cC}$ in Eq. (\ref{eq:n=UcC}). We use polarizer qubits of photon as the flying-qubit. The polarized states can be easily generated and operated in optical apparatus. We choose cold Rydberg atoms as the static qubits to construct nonadiabatic holonomic controlled gates. The flying qubits and the static qubits are entangled in photon-cavity-atom system.

\subsection{Preparing the graph state $\ket{h_3}$}
\begin{figure}[h!]
\centerline{
\includegraphics[width=0.5\textwidth]{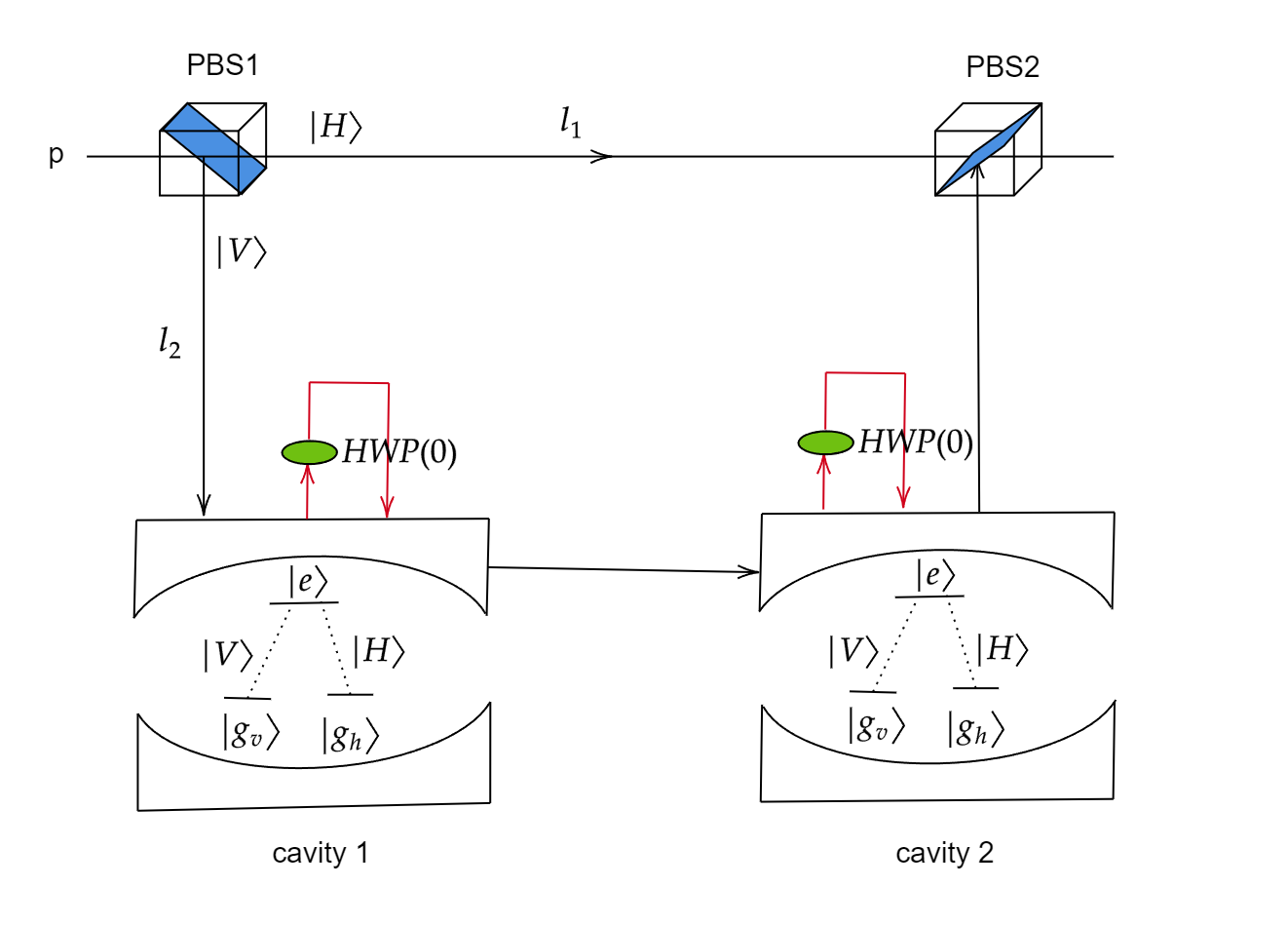}
}\caption{Schematic diagram of CZ gates between Alice, Bob and Charlie in preparing $\ket{h_3}$.}
\label{fig:cavity3}
\end{figure}

Suppose the photon $p$ is controlled by Alice, the initial state is $a_1\ket{H}+a_2\ket{V}$. The horizontal polarized state $\ket{H}$ stands for $\ket{0}$ qubit and the vertical polarized state $\ket{V}$ stands for $\ket{1}$ qubit. The complex coefficients follow the normalization principle $|a_1|^2+|a_2|^2=1$.
When the photon passes through the polarized beam splitter PBS1, $\ket{H}$ passes through and $\ket{V}$ is rejected. They are separated in two different lines $l_1$ and $l_2$ as shown in Fig. \ref{fig:cavity3}.

The target qubits, atom 1 and 2 are trapped in single-sided optical cavity, belonging to Bob and Charlie respectively. They are $\Lambda$-type three level atoms with the ground states $\ket{g_h}$, $\ket{g_v}$ and the excited state $\ket{e}$.
 Suppose the atoms are in states $b_1\ket{g_h}_1+b_2\ket{g_v}_1$ and $c_1\ket{g_h}_2+c_2\ket{g_v}_2$ at the beginning, in which $|b_1|^2+|b_2|^2=1$, $|c_1|^2+|c_2|^2=1$.
The initial state of the hybrid system is
\begin{equation}
\label{eq:n=psi0}
\ket{\psi}_0=(a_1\ket{H}+a_2\ket{V})_p\otimes(b_1\ket{g_h}+b_2\ket{g_v})_1\otimes(c_1\ket{g_h}+c_2\ket{g_v})_2.
\end{equation}
The horizontal
polarized state $\ket{H}$ is transmitted into line $l_1$. It does not interact with atoms, and this part keeps the initial form, $\ket{\psi}_{H}=a_1\ket{H}_{l_1}\otimes(b_1\ket{g_h}+b_2\ket{g_v})_1\otimes(c_1\ket{g_h}+c_2\ket{g_v})_2$.

In line $l_2$, the flying qubits and static qubits are entangled by photon-cavity-atom hybrid gates.
The ground states $\ket{g_h}$ stands for qubit $\ket{1}$ and $\ket{g_v}$ stands for qubit $\ket{0}$. They can be excited to the state $\ket{e}$ by the cavity modes $c_h$ and $c_v$. Suppose $c_h$ and $c_v$ are resonant to the photon's polarized states $\ket{H}$ and $\ket{V}$  respectively.
When the photon is resonant with the cavity-atom system, it feels a hot cavity. The system is performed by a qubit-flip operation. Otherwise if the photon encounters a cold cavity, it is reflected with a phase-flip operation.
The interaction can be summarized as follows \cite{22,3}
\begin{equation}
\label{eq:n=transmit}
\begin{aligned}
  \ket{H}\ket{g_h} & \rightarrow \ket{V}\ket{g_v},\\
  \ket{H}\ket{g_v} & \rightarrow -\ket{H}\ket{g_v},\\
  \ket{V}\ket{g_h} & \rightarrow -\ket{V}\ket{g_h},\\
  \ket{V}\ket{g_v} & \rightarrow \ket{H}\ket{g_h}.
\end{aligned}
\end{equation}
When the photon enters cavity 1 for the first time, the hybrid state in line 2 becomes
\begin{equation}
\label{eq:n=psiV}
\ket{\psi}_{{V1}}=a_2(b_2\ket{H,g_h}_{l_2 1}-b_1\ket{V,g_h}_{l_2 1})\otimes(c_1\ket{g_h}+c_2\ket{g_v})_2.
\end{equation}
After the photon leaves the cavity 1, it passes through $HWP(0)$.

$HWP(\theta)$ is the half-wave plate \cite{2}, with the major axis at an angle $\theta$ to the vertical direction. When a photon passes through them, ignoring the common phase, it experiences a unitary operator $ U^{HWP}(\theta)=\bma	
\cos(2\theta) & \sin(2\theta)\\
\sin(2\theta) & -\cos(2\theta)\\
\ema$.
$HWP(0)$ represents a phase-flip operation $\sigma_z=\ket{H}\bra{H}-\ket{V}\bra{V}$. It transforms the photon state from $b_2\ket{H}_{l_2}-b_1\ket{V}_{l_2}$ to $b_2\ket{H}_{l_2}+b_1\ket{V}_{l_2}$.
In multi-partite case,
Charlie needs to perform Hadamard gate on photon $p_2$ \cite{4}, it can be realized through $HWP(\frac{\pi}{8})$.

 The photon enters cavity 1 again, as shown by the red line in Fig. \ref{fig:cavity3} , Eq. (\ref{eq:n=psiV}) evolves into
\begin{equation}
\label{eq:n=psiV2}
\ket{\psi}_{V2}=a_2(b_2\ket{V,g_v}_{l_2 1}-b_1\ket{V,g_h}_{l_2 1})\otimes(c_1\ket{g_h}+c_2\ket{g_v})_2.
\end{equation}
After these operations, the atom in cavity 1 becomes entangled with photon p. In the subspace spanned by them, the state evolves into
\begin{equation}
\label{eq:n=psiV3}
\ket{\psi}_{p1}=a_1\ket{H}_{l_1}\otimes(b_1\ket{g_h}+b_2\ket{g_v})_1+a_2\ket{V}_{l_2}\otimes(b_2\ket{g_v}-b_1\ket{g_h})_1.
\end{equation}
 These implementations realize a CZ gate between Alice and Bob.

When the photon leaves cavity 1 again, it returns to $a_2\ket{V}_{l_2}$. The implementations are repeated in cavity 2, establishing  a similar CZ gate between Alice and Charlie.
The total final state is
\begin{equation}
\label{eq:n=psit}
\begin{aligned}
\ket{\psi}_{t} & =[a_1\ket{H}_{l_1}\otimes(b_1\ket{g_h}+b_2\ket{g_v})_1\otimes(c_1\ket{g_h}+c_2\ket{g_v})_2\\
 & +a_2\ket{V}_{l_2}\otimes(b_2\ket{g_v}-b_1\ket{g_h})_1\otimes(c_2\ket{g_v}-c_1\ket{g_h})_2].
\end{aligned}
\end{equation}
In line 2, $a_2\ket{V}_{l_2}$ is transmitted back to Alice  and remixed with $a_1\ket{H}$ in PBS2.
So Eq. (\ref{eq:n=psit}) realizes the entangled state $\ket{h_3}$.

\subsection{Entangle c and C}
As described in \cite{24,27}, Raman oscillation can realize local Pauli operations on neutral atoms with less dissipation. In this part, we utilize the Rydberg anti-blockade effect suggested in \cite{28}  to realize a controlled gate between cC.
Nonadiabatic holonomic processing \cite{23,30,29,34} contributes a passage to realize  Eq. (\ref{eq:n=UcC}) directly and efficiently.

Charlie leads atom 2 out of cavity 2, and embeds it in an ODT, together with another atom 3 as qubit C. Qubit C is in an unknown initial state $(d_1\ket{0}+d_2\ket{1})_3$, where $|d_1|^2+|d_2|^2=1$ for normalization.
Both atom 2 and 3 are chosen as alkali metal atoms such as $^{87}Rb$. They are kept at a distance to ensure that the Rydberg blockade scheme works.  Encode two hyperfine ground states as qubit states $\ket{0}$ and $\ket{1}$. They couple with the excited state $\ket{R}$ via laser in  $\Lambda$-type three level processing.

\begin{figure}[h!]
\centerline{
\includegraphics[width=0.5\textwidth]{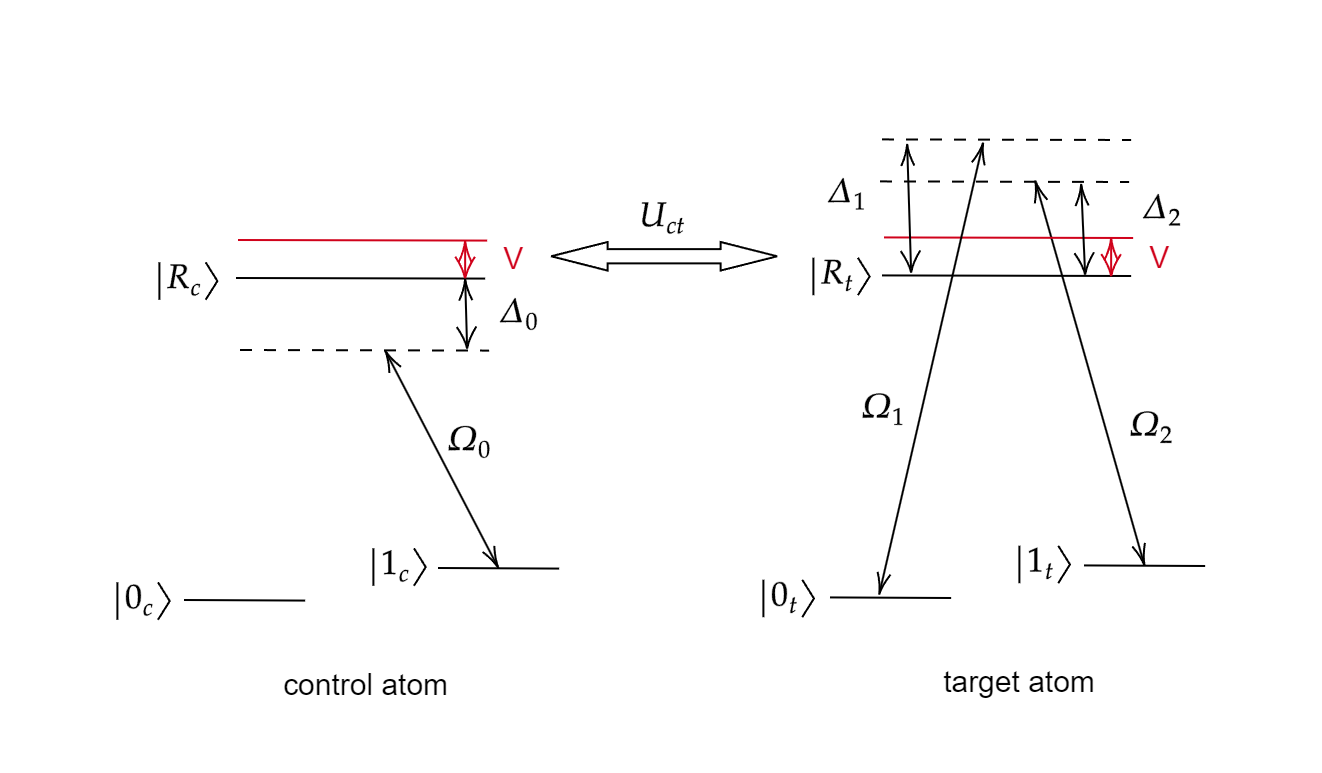}
}\caption{The universal controlled gate is realized by utilizing the Rydberg anti-blockade effect. When control and target atoms are excited to Rydberg states simultaneously in two-photon transitions, the van der Waals interaction between them is $V\equiv U_{ct}=C_6/d^6_{ct}$.}
\label{fig:towqubit gate}
\end{figure}

As shown in Fig. \ref{fig:towqubit gate}, for control qubit c, $\ket{0}_c$ couples to the Rydberg state $\ket{R}_c$ via a laser pulse $\Omega_0$ with red detuning $\Delta_0$. For target qubit C, $\ket{0}_t$ couples to the Rydberg state $\ket{R}_t$ via a laser pulse $\Omega_1$ with blue detuning $\Delta_1$, and $\ket{1}_t$ couples to $\ket{R}_t$ via a laser pulse $\Omega_2$ with blue detuning $\Delta_2$. In interaction picture, the Hamiltonian can be written as

\begin{equation}
\label{eq:n=hami6}
\begin{aligned}
H_I= & \Big[\frac{\Omega_0}{2}e^{-i\Delta_0t}\big(\ket{10}\bra{R0}+\ket{11}\bra{R1}\big)
+\frac{\Omega_1}{2}e^{i\Delta_1t}\big(\ket{00}\bra{0R}+\ket{10}\bra{1R}\big)\\
  & +\frac{\Omega_2}{2}e^{i\Delta_2t}\big(\ket{01}\bra{0R}+\ket{11}\bra{1R}\big)+\frac{\Omega_0}{2}e^{-i(\Delta_0+V)t}\ket{1R}\bra{RR}\\
   &+\frac{\Omega_1}{2}e^{i(\Delta_1-V)t}\ket{R0}\bra{RR}
   +\frac{\Omega_2}{2}e^{i(\Delta_2-V)t}\ket{R1}\bra{RR}+H.c.\Big]+V\ket{RR}\bra{RR}.
\end{aligned}
\end{equation}
 Suppose $\Delta_0\gg\Omega_0$, $\Delta_1\gg\Omega_1$,$\Delta_2\gg\Omega_2$, both atoms are largely detuned. They cannot be excited to Rydberg state individually. All single exciting states can be adiabatically eliminated.
  When the control qubit is in the state $\ket{0}_c$, it is decoupled from $\ket{R}_c$. The target atom remains in its initial state. When the control atom is in $\ket{1}_c$, the interaction of two excited atoms induces the Stark shift V.
 If we set $\Delta_0$ to be close to $\Delta_{1,2}-V$,  the ground states $\ket{10}$ and $\ket{11}$  couple near resonantly with the double-excitation state $\ket{RR}$.
 The two atoms populate between the ground states through a Raman-like process $\ket{10}\leftrightarrow\ket{RR}\leftrightarrow\ket{11}$.
  We can offset the Stark shifts of ground states by auxiliary fields or modified laser pulses \cite{40}.
The effective Hamiltonian can be simplified as
\begin{equation}
\label{eq:n=hami}
H_{eff}=\big(\frac{\Omega_{eff}^{10}}{2}\ket{10}\bra{RR}+\frac{\Omega_{eff}^{11}}{2}\ket{11}\bra{RR}+H.c.\big)+\delta\ket{RR}\bra{RR}.
\end{equation}
The effective frequency is obtained from second-order perturbation theory \cite{28,32},
 \begin{equation}
\label{eq:n=coeffi}
\begin{aligned}
\Omega_{eff}^{10} & =\frac{\Omega_0\Omega_1}{4}e^{i(\Delta_1-\Delta_0-V)}(\frac{1}{\Delta_1}+\frac{1}{\Delta_0+V}-\frac{1}{\Delta_0}
-\frac{1}{\Delta_1-V}),\\
\Omega_{eff}^{11} & =\frac{\Omega_0\Omega_2}{4}e^{i(\Delta_2-\Delta_0-V)}(\frac{1}{\Delta_2}+\frac{1}{\Delta_0+V}-\frac{1}{\Delta_0}
-\frac{1}{\Delta_2-V}),\\
\Delta_{RR} & =\frac{\Omega_0^2}{4(\Delta_0+V)}-\frac{\Omega_1^2}{4(\Delta_1-V)}--\frac{\Omega_2^2}{4(\Delta_2-V)}.
\end{aligned}
\end{equation}
We can set the parameter $\delta\equiv V_0+\Delta_{RR}$ by adjusting the Rydberg blockade factor $V_0\equiv\Delta_{1,2}-(V+\Delta_0)$. If we set the frequency of laser pulses to satisfy $\Delta_{1,2}=V+\Delta_0$, which means $V_0=0$, $\delta=\Delta_{RR}$. It corresponds to a resonant case. If we set $V_0=-\Delta_{RR}$, the Stark shift of double-excited state is absorbed, the effective Hamiltonian can be written as
\begin{equation}
\label{eq:n=hami7}
H_{eff}=\frac{\Omega_{eff}^{10}}{2}\ket{10}\bra{RR}+\frac{\Omega_{eff}^{11}}{2}\ket{11}\bra{RR}+H.c..
\end{equation}

 In the dark-bright basis, suppose the driving fields can be parameterized as  $\frac{\Omega_{eff}^{10}}{2}=\frac{\Omega^{eff}}{2}\sin\frac{\theta}{2}e^{i\varphi}$ and $\frac{\Omega_{eff}^{11}}{2}=-\frac{\Omega^{eff}}{2}\cos\frac{\theta}{2}$.

\begin{equation}
\label{eq:n=dbbasis2}
\begin{aligned}
\ket{B} & =\sin\frac{\theta}{2}e^{i\varphi}\ket{10}-\cos\frac{\theta}{2}\ket{11},\\
\ket{D} & =\cos\frac{\theta}{2}\ket{10}+\sin\frac{\theta}{2}e^{-i\varphi}\ket{11}.
\end{aligned}
\end{equation}
In which, $\ket{D}$ is the dark state of Hamiltonian with zero eigenvalue, $\ket{B}$ is the bright state with
nonzero eigenvalue,
the Hamiltonian becomes
\begin{equation}
\label{eq:n=hami2}
H'_{eff}=\frac{\Omega^{eff}}{2}(\ket{B}\bra{RR}+H.c.).
\end{equation}
The dark state $\ket{D}$  decouples with the excited state. The evolution is reduced to a Rabi oscillation between $\ket{B}$ and $\ket{RR}$ with effective frequency $\Omega^{eff}/2$.
The time evolution operator is $U(\tau)=e^{\int^{\tau}_0\Omega(t')dt'(\ket{RR}\bra{B}+\ket{B}\bra{RR})}$.
A full loop in the Grassmannian is realized when the pulse satisfies $\int^{\tau}_0\Omega(t')dt'=\pi$.
By a cycle evolution  $\ket{B}\mapsto -\ket{B}$, $\ket{D}\mapsto \ket{D}$,
the geometric evolution operator is
\begin{equation}
\label{eq:n=martix}
\begin{aligned}
U(\tau) & =U(\theta,\varphi)=\ket{00}\bra{00}+\ket{01}\bra{01}-\ket{B}\bra{B}+\ket{D}\bra{D}\\
 & =\ket{0}_c\bra{0}\otimes I_C+\ket{1}_c\bra{1}\otimes \sigma_{nC} =\bma	
1 & 0 & 0 & 0\\
0 & 1 & 0 & 0\\
0 & 0 & \cos\theta & \sin\theta e^{i\varphi}\\
0 & 0 & \sin\theta e^{-i\varphi} & -\cos\theta\\
\ema.
\end{aligned}
\end{equation}
 It constructs universal nonadiabatic holonomic controlled gate as Eq. (\ref{eq:n=UcC}) acquiring. The rotate angles $\theta, \varphi$ are adjusted by the frequencies of laser pulses.

  As an example,
 if we choose $\theta=\pi/2$ and $\varphi=\pi$, the operator becomes
 \begin{equation}
\label{eq:n=martix1}
\begin{aligned}
 U(\pi/2,\pi) & =\bma	
1 & 0 & 0 & 0\\
0 & 1 & 0 & 0\\
0 & 0 & 0 & -1\\
0 & 0 & -1 & 0\\
\ema.
\end{aligned}
\end{equation}
 Fig. \ref{fig:fide}(i) shows the populations of the four computational states $\ket{ij}(i,j=0,1)$ and the doubly-excited Rydberg pair state $\ket{RR}$ in some certain case, details are presented in next section. The operator $U(\pi/2,\pi)$ realizes the Raman-like oscillation between $\ket{10}$ and $\ket{11}$. These states exchange their populations with each other in one cycle of evolution.

Besides, Bob  executes the operation $e^{i\alpha\sigma_x}$ on qubit b to realize remote implementations on C. The one qubit gate $e^{i\alpha\sigma_x}=\cos\alpha+i\sin\alpha\sigma_x$ can also be realized by Raman oscillation with an additional geometric phase $\alpha$. Details can be found in \cite{33}.

In five-partite case, we need two photons $p_1$ and $p_2$ belonging to Alice and Charlie respectively. Alice controls the atoms belonging to Bob and David with CZ gate, as Eq. (\ref{eq:n=psit}) described. Charlie controls Eve's atom in the same way. Then David and Eve inject another atoms D and E in their own ODT respectively. D and E are in unknown states and  entangled with d and e  respectively, as Eq. (\ref{eq:n=UcC}) described. In $(2N+1)$-partite case, besides controller Alice, participant $A_j$ sends photon to $A_{N+j}$, $j=3,4,...,N+1$, to construct CZ gate between them.
The protocol shows good extensibility in multi-partite case.

 In this
protocol, atoms are doubly-excited, the dissipation of atoms in Rydberg states decreases the fidelity of the
gate. A dynamical scheme proposed in \cite{28} aims to reduce the population of $\ket{RR}$ by setting $|\delta|\gg|\Omega_{eff}|/2$. The effective Hamiltonian becomes $H_d=-\Omega_d\ket{B}\bra{B}$, where $\Omega_d=\frac{\Omega_{eff}}{2\delta}$ with the gate time $T=\pi/|\Omega_d|$. In this case, as shown in Fig. \ref{fig:fide}(ii), the excitation state $\ket{RR}$ is almost unpopulated, resulting in suppressed dissipation of the atoms. This enables the realization of a more robust gate.

\section{Robustness of CRIO protocol}
\label{sec:fid}
In this section, we discuss the robustness of the protocol. Considering the decay of atoms, cavity damping and Doppler dephasing, we calculate the fidelity and efficiency of the processes discussed in the previous section.

\subsection{Fidelity and efficiency in photon-cavity-atom system}
The actual interactions between the photon and the cavity-atom system are not solely described by Eq. (\ref{eq:n=transmit}). Instead, we must also consider the damping of the cavity and the decay of the two-level atom \cite{22,3},
\begin{equation}
\label{eq:n=transmit2}
\begin{aligned}
  \ket{H}\ket{g_h} & \rightarrow r_{h1}\ket{H}\ket{g_h}+r_{h2}\ket{V}\ket{g_v},\\
  \ket{H}\ket{g_v} &\rightarrow r_0\ket{H}\ket{g_v},\\
  \ket{V}\ket{g_h} &\rightarrow r_0\ket{V}\ket{g_h},\\
  \ket{V}\ket{g_v} & \rightarrow r_{h2}\ket{H}\ket{g_h}+r_{h1}\ket{V}\ket{g_v}.
\end{aligned}
\end{equation}
In which, the reflection coefficients are
\begin{equation}
\label{eq:n=reflection}
\begin{aligned}
 r_{h1} & = \frac{1}{i\omega+\frac{\kappa}{2}}\bigg[\big(i\omega-\frac{\kappa}{2}\big)+\frac{\kappa g^2}{2g^2+\big(i\omega+\frac{\kappa}{2}\big)\big(i\omega+\frac{\gamma}{2}\big)}\bigg],\\
 r_{h2} & =\frac{\kappa g^2}{\big(i\omega+\frac{\kappa}{2}\big)\bigg[2g^2+\big(i\omega+\frac{\kappa}{2}\big)\big(i\omega+\frac{\gamma}{2}\big)\bigg]} ,
\end{aligned}
\end{equation}
where $\kappa$ is the cavity damping rate, $\gamma$ is the decay rate of the atom, $g$ is the coupling strength between the cavity and atom, $\omega$ is the detuning of the photon and the atom-cavity system.

When considering the decoupling of the cavity mode and atom, it describes the interaction between $\ket{H}$ and $\ket{g_v}$ or $\ket{V}$ and $\ket{g_h}$. If $g=0$, the reflection coefficients in Eq. (\ref{eq:n=reflection}) become $r_{h1}=\frac{i\omega-\frac{\kappa}{2}}{i\omega+\frac{\kappa}{2}}, r_{h2}=0$. If $\kappa \gg \gamma, \omega$, we obtain the coefficient $r_0=-1$ in Eq. (\ref{eq:n=transmit2}). The photon is reflected with a phase-flip operation in this case.
If $g^2\gg \kappa\gamma$, the coefficients become $r_{h1}\rightarrow 0, r_{h2}\rightarrow 1$. The states of the system perform qubit-flip operations as described in Eq. (\ref{eq:n=transmit}).

\begin{figure}[h!]
\begin{minipage}{0.33\linewidth}
  \centering
        \centerline{
        \includegraphics[width=1.1\textwidth]{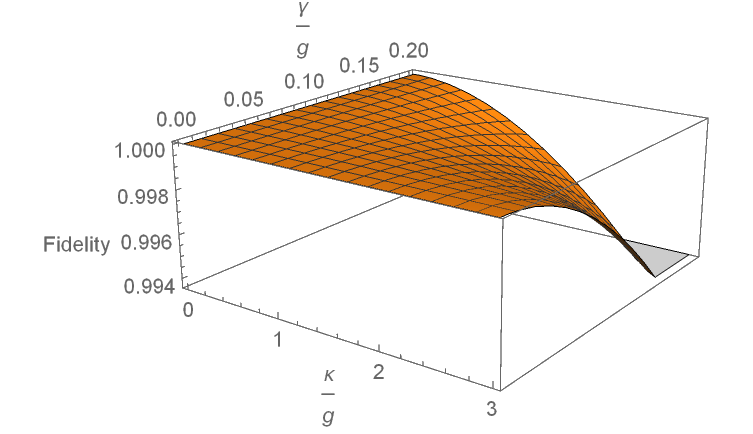}
        }
        \centerline{(i) Fidelity}
    \end{minipage}
    \begin{minipage}{0.33\linewidth}
        \centering
        \centerline{
        \includegraphics[width=1.1\textwidth]{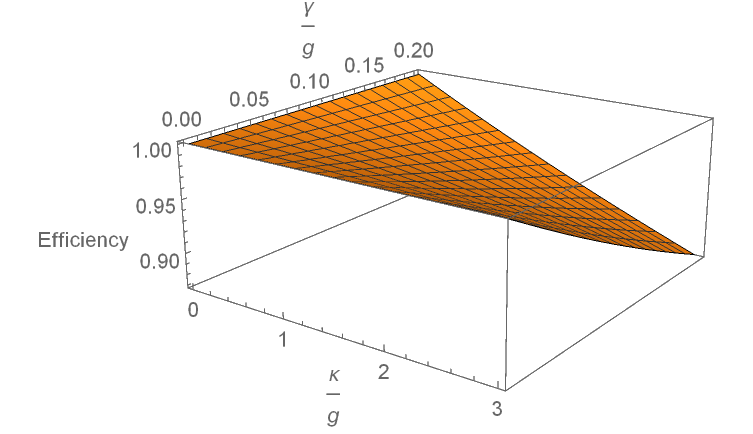}
        }
        \centerline{(ii) Efficiency}
    \end{minipage}
  \caption{The fidelity and efficiency in photon-cavity-atom system.}
\label{fig:fidelity}
\end{figure}

For simplicity, we choose the initial state to be the same as \cite{4}, which means $a_1=a_2=b_1=b_2=c_1=c_2=\frac{1}{\sqrt{2}}$.
So the final state in Eq. (\ref{eq:n=psit}) becomes
\begin{equation}
\label{eq:n=psitideal}
\begin{aligned}
\ket{\psi_{ideal}} & =\frac{1}{2\sqrt{2}}[\ket{H}\otimes(\ket{g_h}+\ket{g_v})_1\otimes(\ket{g_h}+\ket{g_v})_2\\
 & +\ket{V}\otimes(\ket{g_v}-\ket{g_h})_1\otimes(\ket{g_v}-\ket{g_h})_2].
\end{aligned}
\end{equation}
Assuming the detuning $\omega=0$, if we consider the reflection coefficients as a function of $\kappa/g$ and $\gamma/g$ as described in Eq. (\ref{eq:n=reflection}), the final state is
\begin{equation}
\label{eq:n=psitreal}
\begin{aligned}
\ket{\psi_{eff}} & =\frac{1}{2\sqrt{2}}\Big[\ket{H}\otimes(\ket{g_h}+\ket{g_v})_1\otimes(\ket{g_h}+\ket{g_v})_2\\
 & +\ket{Vg_{h1}g_{h2}}-\big(r_{h1}^2-r_{h2}^2\big)\big(\ket{Vg_{h1}g_{v2}}+\ket{Vg_{v1}g_{h2}}\big)+
 \big(r_{h1}^2-r_{h2}^2\big)^2\ket{Vg_{v1}g_{v2}}\Big],
\end{aligned}
\end{equation}
We calculate the inner product of the final state. It follows the normalization principle $|\braket{\psi_{eff}}{\psi_{eff}}/N|^2=1$, where N is the normalization coefficient.
 Define fidelity as $F=|\braket{\psi_{eff}}{\psi_{ideal}}/N|^2$ and the efficiency as $E=|\braket{\psi_{eff}}{\psi_{ideal}}|^2$. During the preparation of $\ket{h_3}$, the fidelity is primarily influenced by the interaction between the photon and the cavity-atom system. Both F and E  are related to the ratios of $\kappa/g$ and $\gamma/g$. As the parameters decrease, their performance can be improved. For example, if $\kappa/g=2$ and $\gamma/g=0.2$, $F=99.54\%, E=90.92\%$, if $\kappa/g=1$ and $\gamma/g=0.2$, $F=99.88\%, E=95.24\%$, and if $\kappa/g=1$ and $\gamma/g=0.1$, $F=99.97\%, E=97.56\%$.
 Details are shown in Fig. \ref{fig:fidelity}.

We can observe that the $\ket{h_3}$ preparation process is robust when $\kappa\gamma$ is smaller than $g^2$.
When we consider long-distance  communication, the polarization mode dispersion (PMD) and chromatic dispersion in optical fiber can reduce the fidelity of the remote CZ gate. There are several optimization methods available, including dispersion compensation and the use of PM fiber. The fiber loss is decreased to 0.36db/km in some PM fibers.

\subsection{Robustness of controlled gate between cC}
 In this part, we calculate the fidelities of nonadiabatic holonomic gate and dynamical gate.
They are influenced by the decay of the atoms and the Doppler dephasing. The evolution of the system cC can be dominated by the master equation,
\begin{equation}
\label{eq:n=mastereq}
\frac{\partial \rho}{\partial t}=-i[\hat{H},\rho]+\mathcal{L}[o]+\mathcal{L}[o'].
\end{equation}
The second and third terms are Lindblad terms, describing the decoherence and dephasing processes of atom respectively.
\begin{equation}
\label{eq:n=lindblad}
\begin{aligned}
\mathcal{L}[o] & =
\sum\limits_{j=\{c,C\}}\sum\limits_{k=\{0,1\}}\Big[\hat{L}_j^k\rho\hat{L}_j^{k\dagger}-
\frac{1}{2}\big(\hat{L}_j^{k\dagger}\hat{L}_j^k\rho+\rho\hat{L}_j^{k\dagger}\hat{L}_j^k\big)\Big],\\
\mathcal{L}[o'] & =\sum\limits_{j=\{c,C\}}\Big[\hat{L'}_j\rho\hat{L'}_j^{\dagger}-
\frac{1}{2}\big(\hat{L'}_j^{\dagger}\hat{L'}_j\rho+\rho\hat{L'}_j^{\dagger}\hat{L'}_j\big)\Big].
\end{aligned}
\end{equation}
The dissipation operator is $\hat{L}_j^k=\sqrt{\Gamma_k}\ket{k}_j\bra{R}$. $\Gamma$ is the decay term linewidth, the lifetime of   Rydberg state is $\tau=1/\Gamma\approx 400\mu s$ \cite{35}.
 Since there are eight ground states of $^{87}Rb$, $\Gamma_{0,1}=\frac{1}{8\tau}$. The dephase operator is $\hat{L'}_j=\sqrt{\kappa_j}(\ket{0}_j\bra{0}+\ket{1}_j\bra{1}-\ket{R}_j\bra{R})$, $\kappa_j$ is close to $\Gamma_{0,1}$ \cite{36}.

We discuss the fidelity of evolution operator $U(\pi/2,\pi)$ in Eq. (\ref{eq:n=martix1}). In density matrix description, the initial-state-specified fidelity is defined as $F=tr(\rho\ket{\Psi_t}\bra{\Psi_t})$, in which $\ket{\Psi_t}=U(\pi/2,\pi)\ket{\Psi_0}$ is the ideal state. Suppose the initial state is $\ket{\Psi_0}=\frac{\ket{0}_c+\sqrt{2}\ket{1}_c}{\sqrt{3}}\otimes\frac{\sqrt{3}\ket{0}_C+\ket{1}_C}{2}$.
We set $\Omega_0=\Omega_1=\Omega_2\equiv\Omega$ for simplicity, and $\Delta_0=10\Omega$ and $\Delta_1=\Delta_2=30\Omega$ to satisfy $\Delta_{0,1,2}\gg \Omega$, $V_0=\Omega/30$. Since $V=\Delta_1-\Delta_0+V_0$, if we choose the distance between two atoms as $d=4.03\mu m$ \cite{28}, the corresponding interaction strength $\frac{V}{2\pi}=200.33MHz$. In this case, $\Omega$ is about $20\pi$ and the gate time $T=\frac{\pi}{|\Omega^{eff}/2|}$. We set $\frac{\delta}{2\pi}=3.36MHz$ in dynamical gate.

\begin{figure}[h!]
\begin{minipage}{0.33\linewidth}
  \centering
        \centerline{
        \includegraphics[width=1.1\textwidth]{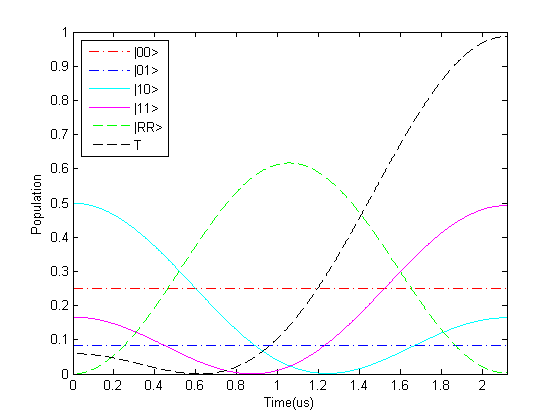}
        }
        \centerline{(i) nonadiabatic holonomic gate}
    \end{minipage}
    \begin{minipage}{0.33\linewidth}
        \centering
        \centerline{
        \includegraphics[width=1.1\textwidth]{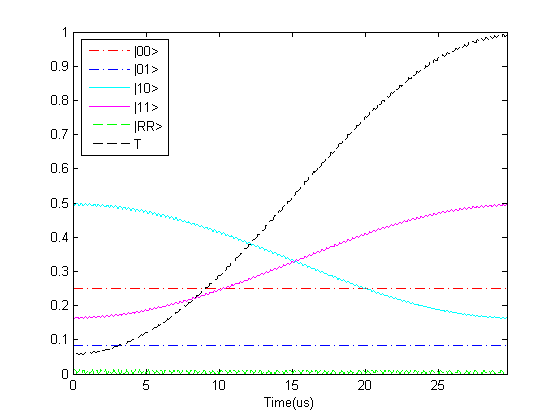}
        }
        \centerline{(ii) dynamical gate}
    \end{minipage}
  \caption{The populations of the states $\ket{ij}(i,j=0,1)$, $\ket{RR}$ and the fidelity of $U(\pi/2,\pi)$ gate.}
\label{fig:fide}
\end{figure}

With these parameters, we plot
the fidelity variation curve in Fig. \ref{fig:fide}. At the end of the gate time, the fidelity reaches $98.65\%$ for the nonadiabatic holonomic gate. It increases to $99.11\%$ for the dynamical gate.

For a universal gate, the rotating angles $\theta$ and $\varphi$ in $U(\tau)$ vary from $0$ to $2\pi$. The average fidelity achieves $99.19\%$ for the nonadiabatic holonomic gate  and  $99.61\%$ for the dynamical gate. Therefore, the universal dynamical controlled gates can consistently produce high fidelity exceeding $99\%$.

 We also examine the impact of the initial states on the fidelity. For arbitrary input states   $\ket{\Psi_0}=\sin\beta_1\ket{00}+\cos\beta_1\big[e^{i\beta_4}\sin\beta_2\ket{01}
 +\cos\beta_2(e^{i\beta_5}\sin\beta_3\ket{10}+e^{i\beta_6}\cos\beta_3\ket{11})\big]$, we average the fidelities for different angels $\beta_i$. For the nonadiabatic holonomic gate $\bar{F}=99.50\%$ and for the dynamical gate $\bar{F}=99.76\%$.

According to these calculations, the protocols we suggested to realize CRIO, exhibit strong robustness against decoherence and dephasing.

\section{Conclusions}
\label{sec:con}
In this article, we presented a set of practical operations to realize CRIO experimentally. We utilized the polarization of photons as flying qubits to transmit remote implementations over long distance. Static qubits were selected as hyperfine structures of neutral atoms due to their stability. The CZ gates between Alice, Bob and Charlie achieved high fidelity in the photon-cavity-atom system. However, the dispersion in long fibers  can not be ignored, PM fiber can  partially eliminate defects. In the future, its performance will be further improved to enhance the reliability of the CRIO protocol.

The local controlled gate in the cC system was achieved using the Rydberg anti-blockade effect. The scheme exhibited strong robustness against atom decay and Doppler dephasing. In this protocol, different laser pulses were utilized to excite the control and target atoms respectively. Therefore, individual addressing of atoms is necessary, which can be achieved using optical tweezers arrays on the neutral atom platform.
These technologies are not only applicable to CRIO,  but also has numerous applications in quantum communication, computation, and information.

%
%
%
%
%

\bibliographystyle{unsrt}

\begin{thebibliography}{1}
\bibitem{6} Yang Liu, Wei-Jun Zhang, Cong Jiang, Jiu-Peng Chen, Chi Zhang, Wen-Xin Pan, Di Ma, Hao Dong, Jia-Min Xiong, Cheng-Jun
Zhang, Hao Li, Rui-Chun Wang, Jun Wu, Teng-Yun Chen, Lixing You, Xiang-Bin Wang, Qiang Zhang and Jian-Wei Pan. Experimental Twin-Field Quantum Key Distribution Over 1000 km Fiber Distance. Phys. Rev. Lett. 130, 210801 (2023).
\bibitem{7} Y.-A. Chen, Q. Zhang, T.-Y. Chen, W.-Q. Cai, S.-K.Liao, J. Zhang, K. Chen, J. Yin, J.-G. Ren, Z. Chen,S.-L. Han, Q. Yu, K. Liang, F. Zhou, X. Yuan, M.-S.Zhao, T.-Y. Wang, X. Jiang, L. Zhang, W.-Y. Liu, Y. Li, Q. Shen, Y. Cao, C.-Y. Lu, R. Shu, J.-Y. Wang, L. Li, N.-L. Liu, F. Xu, X.-B. Wang, C.-Z. Peng, and J.-W. Pan. An integrated space-to-ground quantum communication network over 4,600 kilometres. Nature 589, 214 (2021).
\bibitem{18} Yan-He He, Qiu-Chun Lu, Yue-Ming Liao, Xing-Chen Qin, Jian-Sheng Qin, and Ping Zhou. Bidirectional controlled
23 remote implementation of an arbitrary single qubit unitary operation with epr and cluster states. International Journal
of Theoretical Physics, 54(5):1726-1736, 2014.
\bibitem{19} Li Yu and Kae Nemoto. Implementation of bipartite or remote unitary gates with repeater nodes. Physical Review A,
94:022320, 2016.
\bibitem{20} Neng-Fei Gong, Tie-Jun Wang, and Shohini Ghose. Control power of a high-dimensional controlled nonlocal quantum
computation. Physical Review A, 103:052601, 2021.
\bibitem{21} Nguyen Ba An and Bich Thi Cao. Controlled remote implementation of operators via hyperentanglement. Journal of
Physics A: Mathematical and Theoretical, 55(22), 2022.
\bibitem{4} Xinyu Qiu and Lin Chen. Controlled remote implementation of operations via graph states. Ann. Phys. (Berlin) 2023, 2300320.
\bibitem{5} Benni Reznik, Yakir Aharonov, and Berry Groisman. Remote operations and interactions for systems of arbitrarydimensional hilbert space: State-operator approach. Physical Review A, 65:032312, 2002.
\bibitem{25} Yiru Zhou, Pooja Malik, Florian Fertig, Matthias Bock, Tobias Bauer, Tim van Leent, Wei Zhang, Christoph Becher and Harald Weinfurter. Long-Lived Quantum Memory Enabling Atom-Photon Entanglement over 101 km Telecom Fiber. Arxiv: 2308.08892.
\bibitem{22} Fang-Fang Du, Yi-Ming Wu, and Gang Fan. Refined quantum gates for $\Lambda$-type atom-photon hybrid systems[J]. Advanced Quantum Technologies, 2023.
\bibitem{11} T. Pellizzari. Quantum Networking with Optical Fibres. Phys. Rev. Lett. 79, 5242.
\bibitem{23} Zhu-yao Jin and Jun Jing. Geometric quantum gates via dark paths in Rydberg atoms. Arxiv: 2307.07148.
\bibitem{30}  Emmi Herterich, Erik Sj{\"o}qvist. Single-loop multiple-pulse nonadiabatic holonomic quantum gates. Phys. Rev. A 94, 052310 (2016).
\bibitem{29} Erik Sj{\"o}qvist, D. M. Tong, L. Mauritz Andersson, Bj{\"o}rn Hessmo, Markus Johansson, Kuldip Singh. Non-adiabatic holonomic quantum computation. New J. Phys. 14, 103035 (2012).
\bibitem{34}  Erik Sj{\"o}qvist.  Nonadiabatic holonomic single-qubit gates in off-resonant $\Lambda$ systems. Phys. Lett. A 380, 65 (2016).
\bibitem{37} L. Faoro, J. Siewert, and R. Fazio. NonAbelian holonomies, charge pumping, and quantum computation with josephson junctions. Phys. Rev. Lett. 90, 028301 (2003).
\bibitem{38}  L. M. Duan, J. I. Cirac, and P. Zoller. Geometric manipulation of trapped ions for quantum computation. Science 292, 1695 (2001).
\bibitem{39} P. Solinas, P. Zanardi, N. Zangh\`i, and F. Rossi. Semiconductor-based geometrical quantum gates. Phys. Rev. B 67, 121307(R) (2003).
\bibitem{26} Xiaoling Wu, Xinhui Liang, Yaoqi Tian, Fan Yang, Cheng Chen, Yong-Chun Liu, Meng Khoon Tey, and Li You. A concise review of Rydberg atom based quantum computation and quantum simulation. Chin. Phys. B 30, 020305 (2021).
\bibitem{24} Xiao-Feng Shi. Rydberg Quantum Gates Free from Blockade Error. Phys. Rev. Applied 7, 064017 (2017).
\bibitem{27} Xiao-Feng Shi. Fast, Accurate, and Realizable Two-Qubit Entangling Gates by Quantum Interference
in Detuned Rabi Cycles of Rydberg Atoms. Phys. Rev. Applied 11, 044035 (2019).
\bibitem{41} Kyrylo Simonov, Marcello Caleffi, Jessica Illiano, Angela Sara Cacciapuoti. Universal Quantum Computation via Superposed Orders of Single-Qubit Gates. ArXiv: 2311.13654.
\bibitem{31} Daniel Comparat, Pierre Pillet. Dipole blockade in a cold Rydberg atomic sample.  JOSA B, Vol. 27, Issue 6, pp. A208-A232 (2010).
\bibitem{28}  Jin-Lei Wu, Yan Wang, Jin-Xuan Han, Shi-Lei Su, Yan Xia, Yongyuan Jiang, Jie Song. Unselective ground-state blockade of Rydberg atoms for implementing quantum gates. Frontiers of Physics, 2022, 17(2): 22501.
\bibitem{3} J. H. An, M. Feng, and C. H. Oh. Quantum information processing with a single photon by input-output process
regarding low-Q cavities. Phys. Rev. A 79, 032303 (2009).
\bibitem{2} Englert B G, Kurtsiefer C, Weinfurter H. Universal unitary gate for single-photon two-qubit states. Physical Review A, 2001, 63(3): 32303-32303.
\bibitem{40} Wu J L, Wang Y, Han J X, Feng Y K, Su S L, Xia Y, Jiang Y Y, Song J. One-step implementation of Rydberg-antiblockade SWAP and controlled-SWAP gates with modified robustness[J]. Photonics Research, 2021, 9(5): 814.
\bibitem{32} James D F, Jerke J. Effective Hamiltonian Theory and Its Applications in Quantum Information[J]. Canadian Journal of Physics, 2007, 85(6):625-632.
\bibitem{33} P. Z. Zhao, Xiao-Dan Cui, G. F. Xu, Erik Sj{\"o}qvist, D. M. Tong. Rydberg-atom-based scheme of nonadiabatic geometric quantum computation. Phys. Rev. A 96, 052316 (2017).
\bibitem{35} Beterov I I, Ryabtsev I I, Tretyakov D B, et al. Quasiclassical Calculations of
Blackbody-Radiation-Induced Depopulation Rates and Effective Lifetimes of Rydberg nS , nP, and nD Alkali-Metal Atoms with n$\leq 80$[J]. Physical Review A, 2009, 79:052504.
\bibitem{36} L.-N. Sun, L.-L. Yan, S.-L. Su, and Y. Jia, One-step implementation of time-optimal-control three-qubit nonadiabatic holonomic controlled gates in rydberg atoms, Phys. Rev. Appl. 16, 064040 (2021).
\end{thebibliography}

\end{document}